\documentclass{elsart}

\usepackage{epsfig}

\usepackage{amssymb}

\begin{document}

\begin{frontmatter}



\title{2D and 3D MHD Simulations of Disk Accretion by Rotating  Magnetized Stars:
Search for Variability}



\author{Marina M. Romanova}


\author{Akshay Kulkarni, Min Long, Richard V.E. Lovelace, Justin V. Wick }

\address{Dept. of Astronomy, Cornell University, Ithaca, NY}

\author{Galina V.~Ustyugova, Alexander V.~Koldoba }

\address{Keldysh Institute of Applied Mathematics and Institute of Mathematical Modeling, Russian
Academy of Sciences, Moscow}


\begin{abstract}

We performed 2D and full 3D magnetohydrodynamic simulations of disk
accretion to a rotating star with an aligned or misaligned dipole
magnetic field. We investigated the rotational equilibrium state and
derived from simulations the ratio between two main frequencies:
the spin frequency of the star and the orbital frequency at the inner radius
of the disk. In 3D simulations we observed different features related to the non-axisymmetry of the magnetospheric flow. These
features may be responsible for high-frequency quasi-periodic
oscillations (QPOs). Variability at much lower frequencies may be
connected with restructuring of the magnetic flux threading the
inner regions of the disk. Such variability is specifically strong
at the propeller stage of evolution.
\end{abstract}


\end{frontmatter}

\section{Rotational Equilibrium State and Two Main Frequencies}
\begin{figure}
\begin{center}
\includegraphics*[width=10cm]{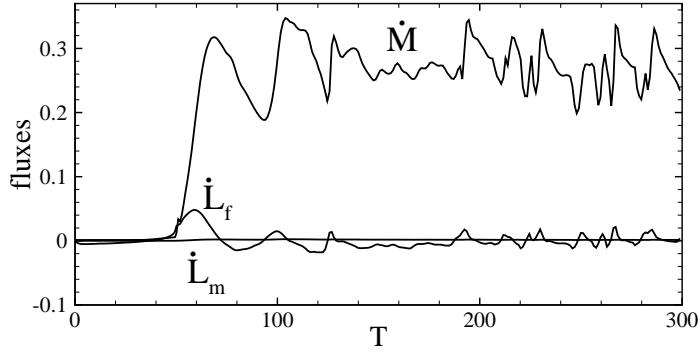}
\end{center}
\caption{Fluxes in the rotational equilibrium state. $\dot M$ is the
matter flux, $\dot L_f$ is the angular momentum flux associated with
the field (it varies in time but is zero on an average), and $\dot
L_m$ is an angular momentum flux associated with matter (it is about
100 times smaller than $\dot L_f$). } \label{fluxes}
\end{figure}
\begin{figure}
\begin{center}
\includegraphics*[width=13cm]{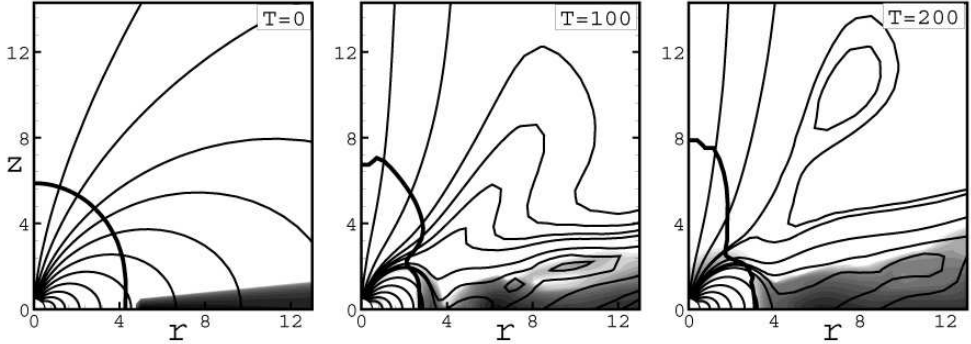}
\end{center}
\caption{Density distribution (background), magnetic field lines
(thin lines) and the $\beta=1$ line (thick line) where the matter
ram pressure equals the magnetic pressure.} \label{Figure 2}
\end{figure}
\begin{figure}
\begin{center}
\includegraphics*[width=11cm]{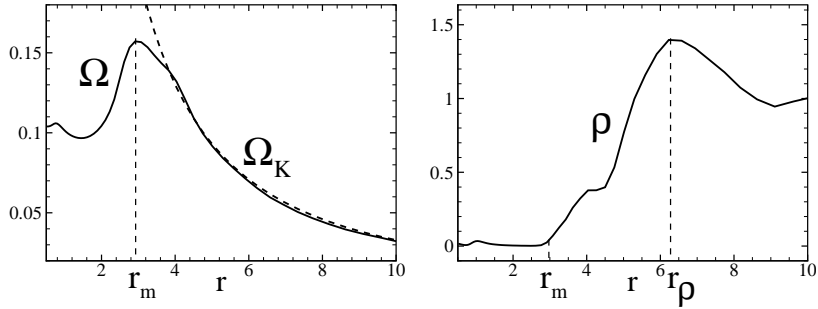}
\end{center}
\caption{{\it Left panel:} Radial distribution of the angular
velocity in the equatorial plane. The dashed line shows the Keplerian
velocity. {\it Right panel:} Density distribution.} \label{Figure 3}
\end{figure}
We first investigated slow viscous disk accretion to a rotating
star with a dipole magnetic field in axisymmetric (2.5D) numerical
simulations. Quiescent initial conditions were developed so that we
were able to observe the accretion for a
long time \cite{rom02}. These simulations have shown that many
predictions of the theories developed in the 1970's (e.g.,
\cite{pri72}--\cite{gho79}) are correct.
  Namely, we observed that
(1) the accreting matter is stopped by the magnetosphere at the
magnetospheric radius $r_m$ ,where the ram pressure $\rho v^2+p$
equals to magnetic pressure $B^2/{8\pi}$, and is then lifted up and
accretes to the star through funnel flows;~ (2) the magnetosphere
influences the structure of the disk to a distance $r_\Psi\approx
(2-4) r_m$; (3) a star may spin-up, spin-down, or may be in the {\it
rotational equilibrium state} depending on the ratio between
magnetospheric radius $r_m$ and corotation radius
$r_{co}=(GM/\Omega_*^2)^{1/3}$, as predicted by Ghosh and Lamb
\cite{gho78}, \cite{gho79}. We searched for the
 rotational equilibrium state and investigated this state in detail
 \cite{lon05}. Namely, we fixed the parameters of the
 disk and magnetic moment of the star, and changed the angular velocity of
 the star, $\Omega_*$. We found angular velocity $\Omega_{eq}$ at which
 the angular momentum
 flux to the star is zero on an average. Figure \ref{fluxes} shows the variation of
 fluxes at the surface of the star for the run with $r_{co}=5$:
 the matter flux $\dot M$ and the angular momentum fluxes $\dot L_m$ and
 $\dot L_f$ associated with
 the matter and the magnetic field respectively.
The flux $\dot L_f$ (which always dominates over $\dot L_m$) varies
but is approximately zero on an average. We found that in the
rotational equilibrium state $r_{co}/r_m\approx 1.4-1.7$.
 Figure 2 shows an example of matter flow in the rotational
 equilibrium state. We see that matter accretes to the star through a funnel flow,
 and that some field lines are closed and others are inflated or radially
 stretched by the accreting matter. Part of the magnetosphere,
 however, is always closed or only partially open, and this
 interaction is important for the spin-up/spin-down of the star.
 The fluxes in Figure 1 vary with time because of inflation
 and reconnection events in the magnetosphere.   We
 observed that in the rotational equilibrium state the rotation of the star
 is locked at some value $\Omega_{eq}\approx k \Omega_m$,
 where $k\approx (1.4-1.6)$ and
$\Omega_m$ is angular velocity of the disk at the magnetospheric
radius. The number $k$ is smaller (larger) at smaller (larger) magnetic moments of the
star $\mu$. These two frequencies, $\nu_*=(2\pi)^{-1} \Omega_*$ and $\nu_m=(2\pi)^{-1}
\Omega_m$ may represent two fundamental frequencies in each binary
system in the equilibrium state. Figure 3 (left panel) shows that
for our typical case, shown in Figure 2, this ratio
$\nu_m:\nu_*\approx 3:2$ (see Figure 3, left panel). This ratio is
reminiscent of the typical ratio between the frequencies of the
kHz QPOs observed from many LMXBs \cite{van00},\cite{wij04}. The
surface of the star may be ``marked" by hot spots associated with
accretion or thermonuclear burning at the surface of the star, while
the inner disk radius, which is approximately equal to
magnetospheric radius $r_m$, may be marked by enhanced density in
the disk or with enhanced reconnection at the boundary between
magnetosphere and the disk.  Matter often accumulates and forms a
larger density peak or, ring,  at the radius $r_{ring}\approx r_m$,
thus giving a ``mark" to this region. However, in some cases, the
density peak is located at larger distances, as is shown in Figure 3
(right panel). The position of this peak varies with the accretion
rate. It is closer to $r_m$ during periods of enhanced accretion. In
case of accretion to a misaligned dipole  the ring typically breaks
into two spiral arms (see next section) which rotate with the
frequency of the inner regions of the disk.

\begin{figure}
\begin{center}
\includegraphics*[width=13cm]{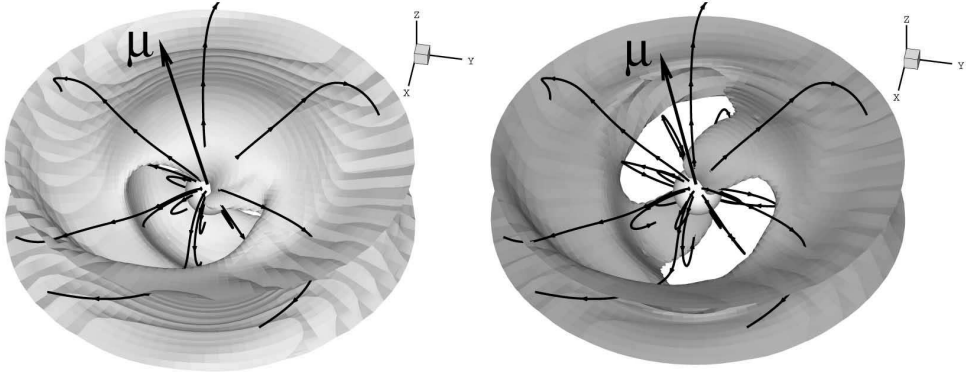}
\end{center}
\caption{Magnetospheric flow is different at different density
levels: matter blankets the entire magnetosphere at $\rho=0.2$ (left
panel), but forms the funnel streams at $\rho=0.4$ (right panel).
The $\Theta=30^\circ$ case is shown.} \label{Figure 4}
\end{figure}

\section{3D Simulations and High-frequency QPOs}

We performed full three-dimensional simulations of the
disk-magnetosphere interaction \cite{rom04a},\cite{kol02},\cite{rom03},
for aligned angular momenta of the star
and the disk, and for different misalignment angles $\Theta$ between
the magnetic and rotational axes. We observed that many features of
the interaction are similar to those in the axisymmetric case.
However, in the 3D case, many new features appear which are
connected with the non-axisymmetry of the flow and which are
important for analysis of high-frequency QPOs.

Simulations have shown that matter typically accretes in two
streams. In cases of relatively high-temperature of the disk
\cite{rom03} or non-stationary accretion, more than two streams may
form.  The shape of the magnetospheric flow strongly depends on the
density. In the same simulation run we see that the largest density
matter  flows in narrow streams, while the lower density matter may
blanket the magnetosphere completely (see Figure 4). So variability
may be related to radiation/obscuration by the magnetospheric
streams. The frequency of this quasi-periodic variability will be
between  $ \nu\sim \nu_* $ and the frequency of the inner region of
the disk, $\nu_m$, which is higher than $\nu_*$ in the rotational
equilibrium state (see Section 1), but may be lower in some other
cases.  If two or several funnel streams form, then the frequency
will be twice or several times higher than that expected in case of
one stream.

\begin{figure}
\begin{center}
\includegraphics*[width=9cm]{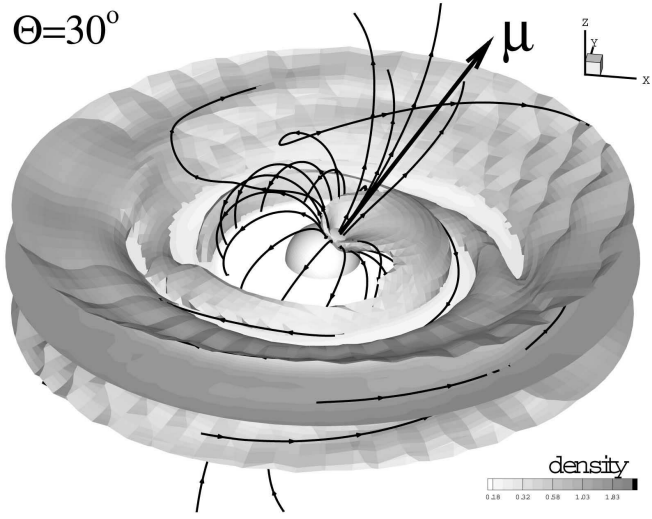}
\end{center}
\caption{Example of magnetospheric flow observed in 3D simulations
for $\Theta=30^\circ$. Inner regions of the disk form a trailing
spiral structure.} \label{Figure 5}
\end{figure}

Other features which may lead to quasi-periodic oscillations are
non-axisymmetric features observed in the inner regions of the disk.
Matter typically forms two-armed spiral waves which may contribute
to quasi-periodic variability (see example in Figure 5).  The
frequency associated with these features is $\nu_{spirals}\lesssim 2
\nu_m$. We should note that in case of the dipole field there are
always two features, from opposite sides of the accretion disk, so
that the frequency is twice as the frequency of the disk.

\begin{figure}
\begin{center}
\includegraphics*[width=13cm]{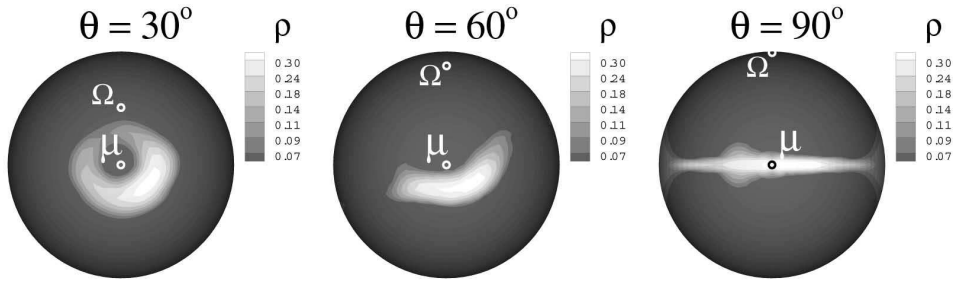}
\end{center}
\caption{Density distribution in the hot spots at the surface of the
star for different $\Theta$ in the relativistic case (from
\cite{kul05}).} \label{Figure 6}
\end{figure}

\begin{figure}
\begin{center}
\includegraphics*[width=13cm]{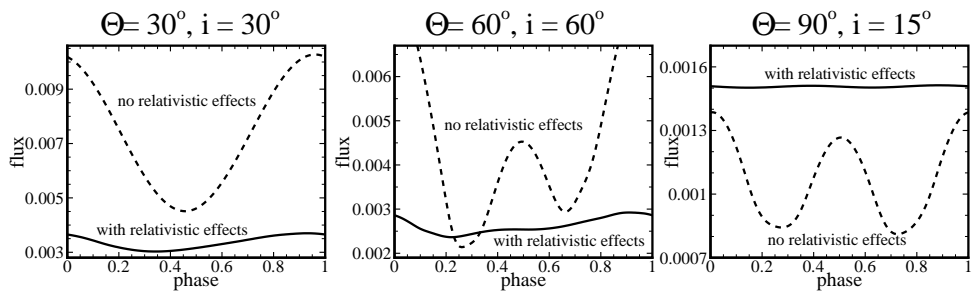}
\end{center}
\caption{Examples of variability curves with relativistic effects
(solid lines) and without relativistic effects (dashed lines) at
different $\Theta$  and inclination angles $i$ (from \cite{kul05}).}
\label{Figure 7}
\end{figure}

Significant variability and quasi-periodic variability may be
associated with the {\it spots} at the surface of the star. The
shape of the spots, and also distributions of density, velocity,
matter and energy fluxes,  reflect those in the funnel streams: at
small misalignment angles $\Theta < 30^\circ$, the spots have the
shape of a bow, while at large misalignment angles $\Theta >
75^\circ$, they have the shape of an elongated bar
\cite{rom04a},\cite {kul05} (see also Figure 6).  It is a common
belief that the hot spots have fixed positions at the surface of the
star. Simulations have shown that the positions of the spots vary
with time. If the disk is relatively cold, or the misalignment angle
is large ($\geq 45^\circ$), then the spots change their shape and
position only slightly, so that their frequency slightly varies
around stellar frequency $\nu\sim \nu_*$. However, if the disk is
relatively hot and the misalignment angle is small ($\leq
30^\circ$), then the funnel streams and spots may rotate faster or
slower than the star, because one of foot-points is dragged by the
disk, which may rotate faster or slower than the star. This may lead
to significant departure of the associated frequency $\nu_{spot}$
from the $\nu_*$. The expected frequency $\nu _{spot}$ is in between
$\nu_*$ and $\nu_m$. Temporary events of precession of the funnel
streams (and spots) are expected in periods of non-stationary
accretion.

The light curves from the hot spots are different for different
misalignment angles $\Theta$ and inclination angles $i$. To
investigate these dependencies, we fixed hot spots at the surface of
the star at some moment corresponding to quasi-equilibrium,  and
then investigated light curves for different $\Theta$ and $i$
assuming that all kinetic energy of the flow is converted to
black-body radiation \cite{rom04a}. We observed that the amplitude
of variability may be quite large and either one or two peaks can be
observed depending on $\Theta$ and $i$. Typically, two peaks are
observed at larger $\Theta$ and $i$. However, in accretion to
neutron stars, general relativistic, Doppler, and other effects are
important (e.g., \cite{pech83},\cite{pg03}). Typically, the GR
effect dominates unless the star rotates extremely fast. We
performed 3D simulations taking into account GR effects
(approximated by the Paczy\'nski-Wiita potential) and other effects
for neutron stars with parameters close to those in millisecond
pulsars. We observed that the shapes of the spots are very similar
to those in the non-relativistic case \cite{kul05} (see Figure 6).
We also observed that light bending effects strongly decrease the
amplitude of the variability curves \cite{kul05} (see Figure 7),
agreeing with earlier predictions (see, e.g. \cite{bel02}). In
extreme cases, like $\Theta=90^\circ$ and $i=15^\circ$, both spots
are observed at the same time, so that the amplitude of variability
becomes very small (Figure 7, right panel; see also \cite{bel02}).

These two main features associated with (1) spots at the star and
(2) spirals at the inner radius of the disk  may lead to
high-frequency quasi-periodic oscillations. Simulations have shown
 that in case of the dipole magnetic field there will be {\it two
symmetric features} (two spirals and two hot spots) and this will
double the frequency. We should note that there is another important
frequency, a beat frequency: $\nu_b=\nu_m-\nu_*$ which is taken into
account in the Beat-Frequency and related models (see \cite{alp85},
\cite{lam85}). This frequency will correspond to the lower frequency
oscillations. If magnetic field of the star is dynamically
unimportant then the Sonic-Point model may be applicable for
analysis of QPOs (\cite{mil98}). However we did not model cases with
such a weak magnetic field.

\begin{figure}
\begin{center}
\includegraphics*[width=14cm]{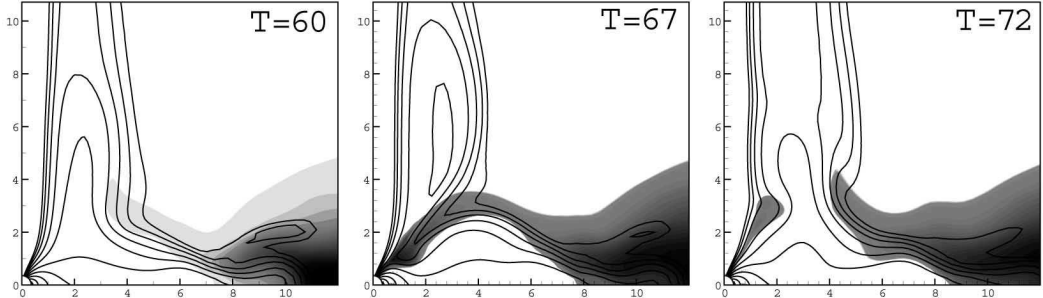}
\end{center}
\caption{Example of accretion at the weak propeller stage: matter
accumulates and accretes quasi-periodically to the star, forcing
magnetic field lines in the magnetic ``tower" to reconnect (from
\cite{rom04b}).} \label{Figure 8}
\end{figure}

\section{``Propeller" Stage and Low-frequency QPOs}

Simulations often show variability of fluxes at much longer
time-scales (see, e.g.,  figure 1). This variability is connected
with restructuring of the external magnetic field lines in the
corona as a result of the disk-magnetosphere interaction. Such
variability appears at different rotation rates of the star, but it
is the strongest at the ``propeller" regime of evolution.

We studied disk accretion to a star in the propeller regime by
axisymmetric MHD simulations. In the propeller regime the
centrifugal force at the external boundary of the rotating
magnetosphere is larger than the gravitational force, so that
accreting matter has a tendency to be flung away (e.g., \cite{ill75}
-- \cite{eks05}). We observed that there are two qualitatively
different regimes:   ``weak" propeller regime
 (no outflows) or ``strong'' propeller regime (with outflows). In both
 cases the star strongly spins-down due
 to the interaction of the magnetosphere with the inner regions of
 the disk.

The weak propeller  occurs when the transport coefficients in the
disk  (viscosity and diffusivity) are relatively small. Both
viscosity and diffusivity were incorporated to the code and were
considered in the simplified $\alpha-$ approach \cite{sha73}. The
weak propeller realizes  at relatively small transport coefficients,
$\alpha_{\rm vis} < 0.2$ and $\alpha_{\rm dif} < 0.2$. We observed
that magnetic field lines inflate, forming a magnetic tower (see
also \cite{kat04}), that the closed magnetosphere expands in the
radial direction, and matter accumulates near the magnetosphere and
is blocked by the inflated magnetic flux from accretion. At some
point, the accreting matter moves forward, forces magnetic field
lines in the magnetosphere to reconnect and accretes to the star.
Later, the magnetosphere expands again and the process repeats
\cite{rom04b} (Figure 8). Such quasi-periodic accretion events have
lower frequencies $\nu_{wp}$ at larger magnetic moments $\mu$ of the
star and larger $\Omega_*$. The typical frequency of the
oscillations lies in the range $\nu_{wp}\approx 0.01 \nu_*$ to
$\nu_{wp}\approx 0.2 \nu_*$ depending on parameters of the star and
the disk.

The strong propeller (with outflows) was obtained when the viscosity
and diffusivity in the disk were larger, $\alpha_{\rm vis} \geq 0.2$
and $\alpha_{\rm dif} \geq 0.2$.
 We observed that matter is ejected in the form of conical,
wide-angle  jets as a result of  magneto-centrifugal forces
\cite{bla82},\cite{ust99}. The process is quasi-periodic and we
observed many periods of accretion and outflows. An example of
outflows is shown in Figure 9. Important factors which led to
propeller-driven outflows are the larger viscosity and diffusivity
in the disk, which helped to transport momentum from the
magnetosphere to the disk. Also, the enhanced viscosity helped the
disk matter penetrate deeper into the rapidly rotating magnetosphere
of the propeller. Both accretion and outflows occur
quasi-periodically, which is connected with periods of (1) inward
viscous accretion; (2) matter diffusion through the outer regions of
magnetosphere and angular momentum transport from the magnetosphere
to the disk; (3) accretion to the star and outburst to the jet; (4)
outward motion of the disk (see also \cite{kat04}, \cite{goo97}).
The span of frequencies of quasi-periodic oscillations is similar to
that in the weak propeller case, $\nu_{sp}\sim (0.01-0.2) \nu_*$.
Figure 10 shows an example of the quasi-variable matter fluxes with
a dominant low oscillation frequency $\nu_{sp}\approx (0.015-0.02)
\nu_*$. We also noticed that at even higher viscosity, $\alpha_{\rm
vis}=0.6$, the quasi-periodic oscillations became very ordered (see
Figure 11). Their frequency was observed to drift slowly from
$\nu_{sp}\approx 0.1 \nu_*$ at the beginning of the simulation to
$\nu_{sp}\approx 0.2 \nu_*$ at the end of the simulation after
$>1000$ rotations of the star, because the disk moved closer to the
star. Similar oscillations were observed in a test run with even
higher viscosity \cite{rom05}.

We should note that the low-frequency variability is also observed in
case of slowly rotating stars as a result of the restructuring of
the magnetic field lines threading the inner regions of the disk
(see Figure 1).

\begin{figure}
\begin{center}
\includegraphics*[width=14cm]{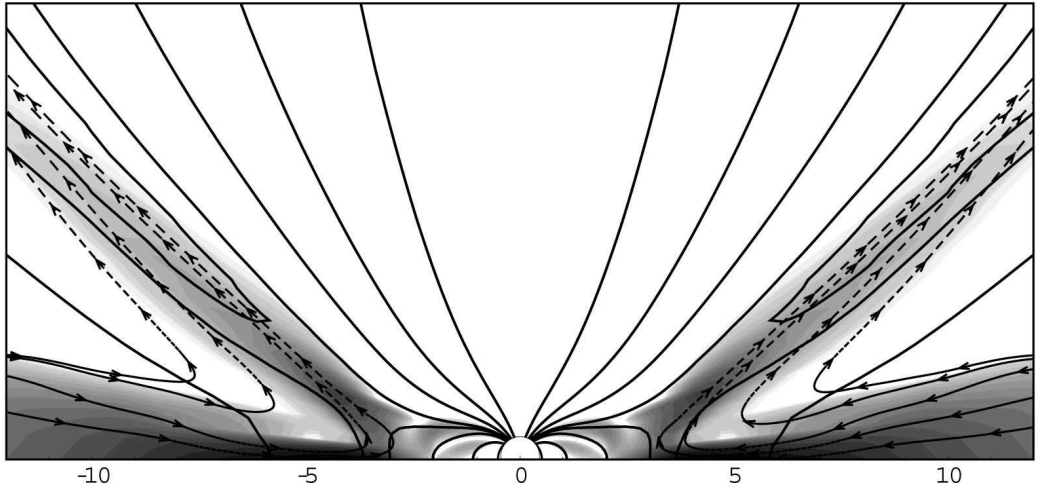}
\end{center}
\caption{Example of outflows at the ``propeller" stage. Background
and dashed lines show the angular momentum flux carried by matter. Solid
lines are magnetic field lines.} \label{Figure 9}
\end{figure}

\section{Conclusions}

Analysis of the disk-magnetosphere interaction in axisymmetric and
full 3D MHD simulations have shown that: (1) In the rotational
equilibrium state, the ratio between frequencies of the inner disk
and of the star is $\nu_m : \nu_*\approx (1.4 - 1.6)$ for
magnetospheric radii $r_m\approx (3-6) R_*$. This ratio may possibly
be larger for larger $r_m$; (2) Non-axisymmetric features in the
inner regions of the disk (typically, two-armed spirals,
non-axisymmetry of the central density ring may lead to QPO
oscillations with $\nu \sim 2 \nu_m$; (3) At small misalignment
angles $\Theta$ the funnels may precess, and thus may rotate
faster/slower than the star; the frequency will vary between $\nu_*$
and $\nu_m$ or twice these values (depending on the number of
spots); precession of funnels is expected during periods of
non-stationary accretion; (4) In all cases the hot spots slightly
change their position and shape which may lead to low-amplitude QPOs
with frequency $\nu_{spots}\approx \nu_*$; (5) Low-frequency
quasi-periodic oscillations were observed at the ``propeller" stage,
with dominant frequency in the range $\nu_p \leq (0.2 - 0.01)
\nu_m$.

\smallskip

\begin{figure}
\begin{center}
\includegraphics*[width=10cm]{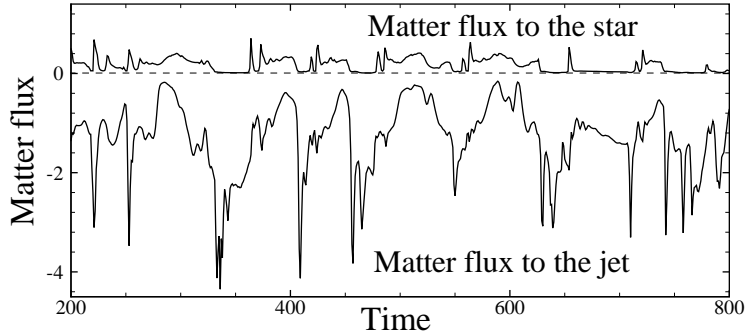}
\end{center}
\caption{Matter fluxes to the star and to the jet when $\alpha_{\rm
vis}=0.3$ and $\alpha_{\rm dif}=0.2$. Time is measured in periods of
Keplerian rotation at $r=1$. } \label{Figure 10}
\end{figure}

\begin{figure}
\begin{center}
\includegraphics*[width=10cm]{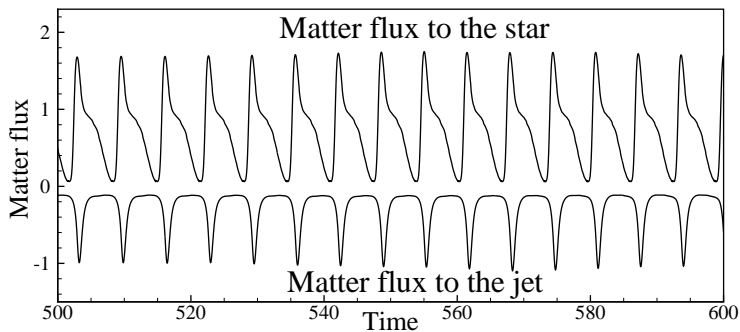}
\end{center}
\caption{Matter fluxes at $\alpha_{\rm vis}=0.6$ and $\alpha_{\rm
dif}=0.2$. Time is measured in periods of Keplerian rotation at
$r=1$.} \label{Figure 11}
\end{figure}

\noindent {\bf Acknowledgments}.~ We are grateful to Prof. Ghosh for
organizing such an excellent meeting and to Prof. F. Lamb and Prof.
Poutanen for useful discussions. This work was supported in part by
NASA grant NNG05GG77G,  and by NSF grants AST-0307817 and
AST-0507760. AVK and GVU were partially supported by grant RFBR
06-02-16548.

\end{document}